# Multiple Neuronal Specializations Elicited By Socially Driven Recognition Of Food Odors

Viktor Plusnin[1,2], Nadezhda Khoroshavkina[3], Ali Abonakour[4,5], Arsenii Onuchin[6,7], Viktoriya Manyukhina[8], Nadezhda Titova[9], Artem Kirsanov[1], Vasiliy Solodovnikov[10], Nikita Pospelov[1,2], and Vladimir Sotskov[1,2,11,12]

[1]Institute for Advanced Brain Studies, Lomonosov Moscow State University, Moscow, Russia

[2]Laboratory of Neuronal Intelligence, Lomonosov Moscow State University, Moscow, Russia

[3]International Laboratory of Cluster Geometry, National Research University Higher School of Economics, Moscow, Russia

[4]Institute of Higher Nervous Activity and Neurophysiology RAS, Moscow, Russia

[5]Moscow Institute of Physics and Technology, Dolgoprudny, Russia

[6]Skolkovo Institute of Science and Technology, Moscow, Russia

[7]Laboratory of Complex Networks, Center for Neurophysics and Neuromorphic Technologies, Moscow, Russia

[8]Center for Neurocognitive Research (MEG Center), Moscow State University of Psychology and Education, Moscow, Russia

[9]Institute of Cognitive Neuroscience, National Research University Higher School of Economics, Moscow, Russia

[10]Laboratory for Neurobiology of Memory, P.K.Anokhin Institute of Normal Physiology, Moscow, Russia

[11]Center of Interdisciplinary Research in Biology, Collège de France, Paris, France

[12]Institute of Biology, École Normale Supérieure, Paris, France

## ABSTRACT

This study investigates the dynamics of non-spatial specializations in hippocampal place cells during exposure to novel environments. Hippocampal place cells, known for their role in spatial mapping, exhibit multi-modal responses to sensory cues. The research focuses on understanding how these cells adapt their specialization in response to novel stimuli, specifically examining non-spatial determinants such as odors and social interactions. Using a social-driven food odor recognition model in mice, the study records CA1 hippocampal neuron activity through miniscope imaging. The experimental design involves demonstrations of novel odors to mice, followed by observation sessions with food options. The analysis employs deep neural network tools for behavior tracking and the custom-developed INTENS software package for identifying neural specializations. Results indicate multiple specializations, particularly those related to odor, with differences observed between training and testing sessions. The findings suggest a temporal aspect to the formation of these specializations in novel conditions, necessitating further investigation for precise tracking.

# INTRODUCTION

Hippocampal place cells have been implicated in a wide range of cognitive tasks, including but not limited to the formation of a cognitive map of space during exposure to novel environments [1]. Prior studies have revealed that the hippocampal place cells are a highly multimodal class of neurons, receiving information from a wide variety of sensory modalities and accurately localizing their firing pattern to restricted regions of an environment [2]. Although place fields undeniably occur in specific regions of space, these locations might be specified by some mixture of cues such as odors, visual cues, touch, and particular behavioral parameters. This process is widely assumed to involve interconnection between hippocampus and neocortical regions responsible for representing sensory information associated with the original event [1]. There are two main categories of non-spatial determinants of place cell activity [2,3]. The first includes stimuli that are not strictly spatial but contribute to the overall representation of a place, such as color, texture, odor, and the expectations or intentions of the animal. These stable stimuli help the animal understand its environment and anticipate future events or actions. The second category consists of discrete events like tone-conditioned stimuli [4], reinforcement learning [5], fear conditioning and odor cues [6]. However, despite its fundamental importance, the emergence and dynamics of non-spatial specializations of place cells are still largely undescribed. In particular, it is unclear how such multi-specialized place cells, which have selective activity both in relation to certain place fields and in relation to certain behavioral parameters, change their specializations in the face of novelty, when animals have to form a new behavior program.

To address this problem, we recorded the activity of CA1 hippocampal neurons in mice in the model of social driven food odor recognition, adapted from [7]. This model makes it possible to observe the wide variety of behavior in laboratory conditions and provides the balance between easiness of behavior readout and naturality of behavior. To record neural activity we used miniscopes, which are head-mounted devices designed for calcium imaging in freely behaving rodents [8].

## MATERIALS AND METHODS

### Animals and surgery

Eight adult C57BL/6 mice were taken for this study and prepared for miniscopic imaging. Surgery protocols were described in many details in our previous works [9,10].

First, a AAV-DJ-CAG-GCaMP6s viral particles were injected to the field CA1 of the hippocampus of anesthetized mice (Bregma: −1.9 mm AP, −1.4 mm ML, 1.25 mm DV, see Fig.1A) through a glass micropipette at a rate of 100 nl/min. Two weeks later mice were anesthetized again and 1.0 mm diameter GRIN lens probe (Inscopix Inc.) was lowered slowly to a depth of 1.1 mm after the partial aspiration of the cortical tissue. Another two weeks later the animals were checked for a fluorescent calcium signal under anesthesia. The mice were fixed in the stereotaxis, and the nVista HD miniature microscope (Inscopix Inc.) was lowered upon the GRIN lens probe and the optimal field of view was chosen. Then a baseplate for chronic imaging was affixed to the skull surface with dental acrylic. After a one-week recovery, the Nvista HD miniscope was attached to awake mice and fluorescent signal was checked.

### Experimental setup and imaging sessions

The animals were divided into 2 equal groups, demonstrators and observers. Prior to the imaging sessions, all mice were provided with a neutral odorless food to adjust and adapt to receiving food from custom made cups-feeders. Then mice-demonstrators were presented food with a novel odor (cocoa or cinnamon). In the first imaging session, mice-demonstrators were paired with mice-observers with the attached nVista HD miniscope and were placed in pairs for 60 minutes into a rectangle arena (20x30 cm). Mice were allowed to explore the environment in arbitrary directions and to interact with each other. The second imaging session was conducted 24 h apart. During this session mice-observers with attached Vista HD miniscope were placed in the arena same to the first session and were allowed to eat food from two cups, where the food with the novel odor, presented earlier to the mice-demonstrators and the food with the control odor (cinnamon or cocoa) was placed. In both imaging sessions, the neural activity was recorded at 20 frames per second with NVista HD v 2.0 miniscope, and the video of mice behavior was captured with a Sony HDR CX-405 (Sony Corp.) camera at 25 frames per second.

Data processing

Miniscopic calcium video recordings were processed with a custom Python pipeline [11] for the detection and inspection of active neuronal units. This pipeline is based on NoRMCorre routine [12] for motion correction and CNMF [13] for neural unit detection. The manual inspection was used to assess the quality of obtained neural units. Matching of neurons between sessions was performed by CellReg routine [14], based on spatial footprints of the detected neural units.

Behavior video recordings were processed using the deep neural network tool [15]. 200 frames were taken from each recording for training of the neural network. Then the ears, the center of the head, the center of the body, the base of the tail, the front and hind limbs, and the nose of the mice were marked on each frame. As a result of training the neural network, a model was obtained capable of recognizing the target points of an animal in each frame of the animal's behavior recording.

Then, discrete behavior states were annotated using our custom Matlab pipeline [16], and continuous kinematic features were constructed using the points recognized by the model.

Detection of neural specializations

We used the INTENS open software package for identifying cognitive specializations of neurons in various conditions [17]. This pipeline is based on calculation of gaussian copula mutual information (GCMI) and searching for statistically significant information connections, including non-linear ones, between the calcium signal and the behavioral variables, both discrete and continuous, to automatically annotate "neuron-behavioral variable" pairs, which are tested for statistical significance by random shuffling of time series. The pairs with p-value more than 0.05 were discarded and the remaining pairs were annotated as statistically significant and the neurons were considered specialized for a given behavior feature.

**RESULTS & DISCUSSION**

We recorded the natural activity and behavior of freely moving mice in the model of socially driven recognition of food odors. In this model, we divided mice into 2 groups, demonstrators and observers. The experiment consisted of two imaging sessions, interaction and test sessions. In the interaction session, mice-observers were allowed to interact with demonstrators, which were pre-trained to eat food with an odor, novel for the observers. In the test session, mice-observers were allowed to eat from cups with two odors, the one that was presented to the demonstrators and the control one. The scheme of the experiment design can be seen on fig. 1.

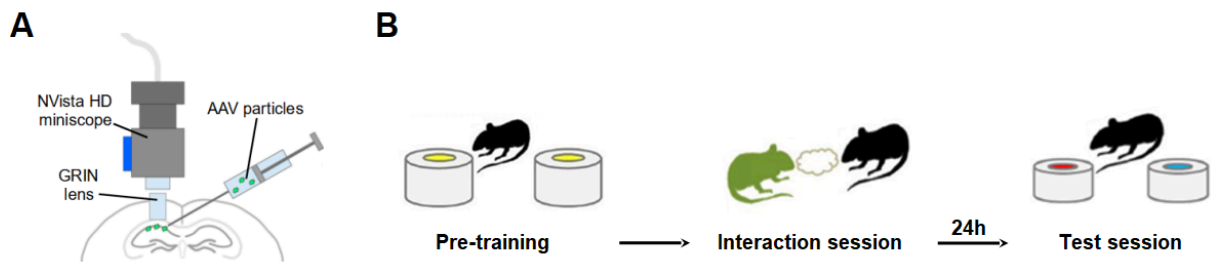

**Figure 1.** Experiment design. A scheme of viral injection, GRIN lens implantation, and miniscope imaging (**A**) and the timeline of experiment (**B**).

During these imaging sessions, we successfully imaged 950 and 919 neurons in 4 mice in the interaction session and in the test session respectively, and matched 366 neurons between the imaging sessions. We applied our newest INTENS method to the data and discovered 713 and 1019 specialized neurons in the interaction and in the test sessions, respectively, and found that multiple specializations occurred in 5 cases in the interaction session and in 79 cases in the test session. The example of such specializations can be seen in Fig. 2.

The presence of such multiple specializations is consistent with previous studies [3,5,6]. However, small amounts of detected specializations are insufficient for statistical analysis and more experimental data are needed to perform quantitative analysis of the emergence and evolution of such specializations.

Mainly, multiple specializations connected with odor were observed at the testing session. This may imply that some time is needed for their formation in the novel conditions like that used in our experimental model.

Taken together, these results might serve as a starting point for further experiments aiming to track the development of complex neuronal specializations in conditions of free behavior with social interaction.

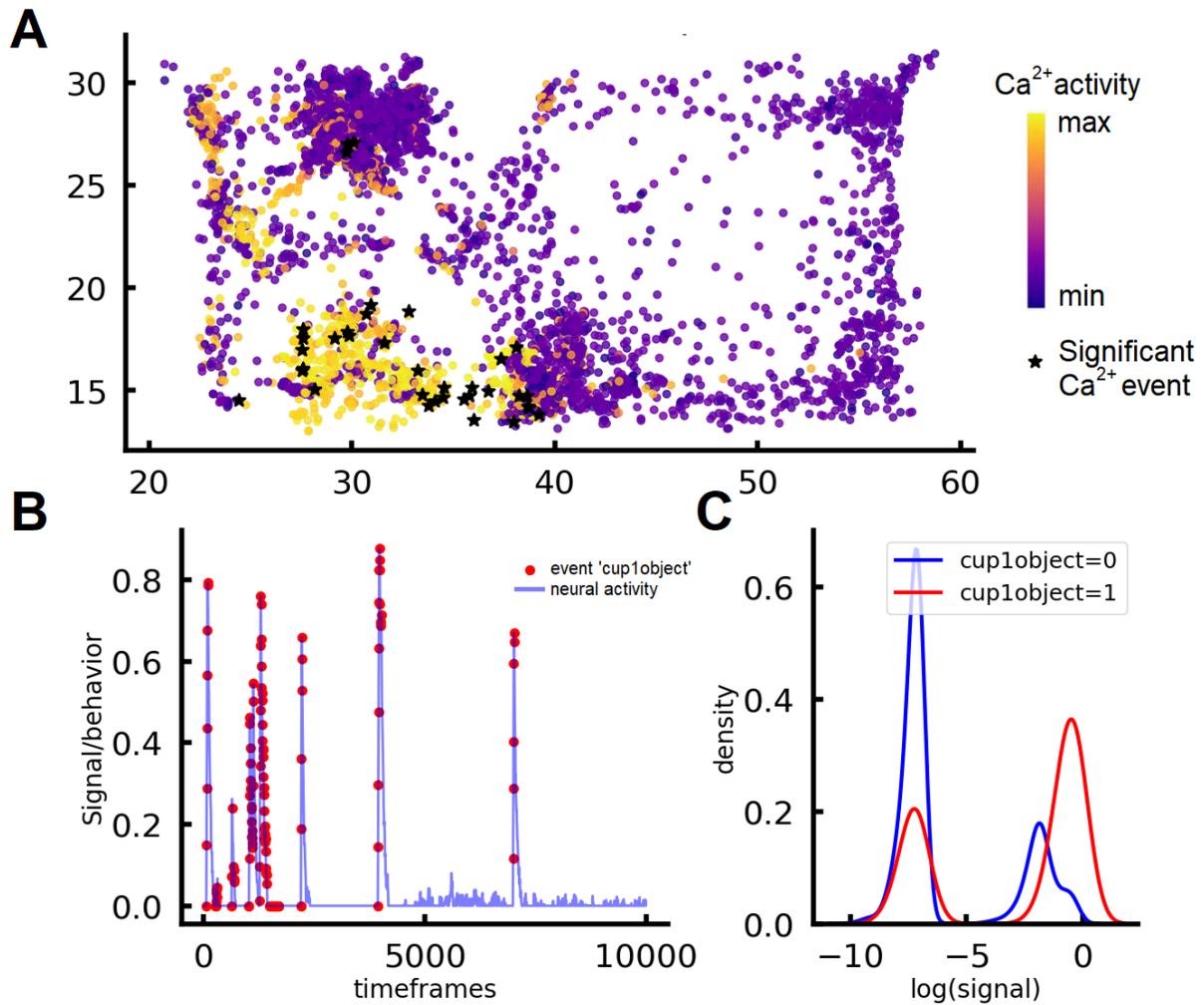

**Figure 2.** An example of multiple specializations of a neuron. **A.** Spatial map of calcium activity of a neuron which demonstrates the properties of a place cell, i.e., specialization on the feature "(x & y)". **B.** Neural activity of the same neuron and discrete behavioral feature "approaching to the cup1 object", where cup1 is a feeder with a demonstrated odor. **C.** Distribution of points (frames) by the neural activity occurred at that points, points when the feature was present and the ones when it was not were plotted separately. These plots show that higher neural activity is associated with the presence of feature.

### Acknowledgments

This study was supported by the IDEAS Scientific Foundation.

### Ethical statement

All methods for animal care and all experimental protocols were approved by the National Research Center "Kurchatov Institute" Committee on Animal Care (NG-1/109PR of 13 February 2020) and were in accordance with the Russian

Federation Order Requirements N 267 M3 and the National Institutes of Health Guide for the Care and Use of Laboratory Animals.